\documentclass[12pt,titlepage]{article}
\usepackage{epsfig}
\usepackage{graphicx}
\usepackage{amsmath}
\usepackage{color}

\def\beq{\begin{equation}}
\def\eeq{\end{equation}}
\def\bea{\begin{eqnarray}}
\def\eea{\end{eqnarray}}

\def\E{{\cal E}}

\def\a{\bar a}
\def\l{\bar \lambda}

\def\V{{\varpi}}
\def\cap{{\cal C}_S}

\begin{document}

\title{The intrinsic curvature of thermodynamic potentials for black holes with critical points}
\author{{Brian P. Dolan}\\
{\small Dept. of Mathematical Physics, 
Maynooth University, Maynooth, Ireland}\\
{\small Dublin Institute for Advanced Studies, 10 Burlington Rd, Dublin, Ireland}\\
{\small Email:} {\tt bdolan@thphys.nuim.ie}
}

\maketitle

\abstract{The geometry of thermodynamic state space is studied for
asymptotically anti-de Sitter black holes in $D$-dimensional space times.
Convexity of thermodynamic potentials and the analytic structure of the response functions is analysed.  The thermodynamic potentials can be used to define a metric on the space of thermodynamic variables and two commonly used such metrics are the Weinhold metric, derived from the internal energy, and the Ruppeiner metric,
derived from the entropy.
The intrinsic curvature of these metrics is calculated for charged and for rotating black holes and it is shown that the curvature diverges when heat capacities diverge but,
contrary to general expectations, the singularities in the Ricci scalars
do not reflect the critical behaviour. 

When a cosmological constant is included as a state space variable it can be interpreted as a pressure and the thermodynamically conjugate variable as a thermodynamic volume.  The geometry of the resulting extended thermodynamic state space is also studied, in the context of rotating black holes, and there are curvature singularities when the heat capacity at constant angular velocity diverges and when the black hole is incompressible.  Again the critical behaviour is not visible in the singularities of the thermodynamic Ricci scalar.}

\bigskip
\leftline{PACS nos: 04.50.Gh; 04.70.Dy; 05.70.-a \hfill DIAS-STP-2015-01}

\section{Introduction}

Thermodynamics is a very general yet remarkably powerful tool
for understanding the physical properties of a very wide range of phenomena.
Its generality lies in the fact that the formalism sidesteps direct questions
about the nature of the underlying microscopic physics. This is an obvious
advantage when studying systems whose microscopic physics is not yet well
understood, such as the physics of quantum gravity. The quantum nature of
black holes is thus a perfect arena for thermodynamics 
to be used as an early stage investigative tool for gaining insight into
the underlying microscopic physics.

The aim of the present study is to elucidate the geometry of black hole thermodynamic potentials.
For a gas thermodynamic stability demands that the thermal
energy be a convex function of the entropy $S$ and the volume $V$, from which the well known relation
$C_P>C_V$ follows. The isothermal compressibility $\kappa_T$ and the isentropic
compressibility  $\kappa_S$ are related to the heat capacities by $C_P \kappa_S=C_V \kappa_T$, implying the reciprocal relation $\kappa_T > \kappa_S$.
Schwarzschild black holes famously have negative heat capacity \cite{Hawking}
but
stability can be achieved in asymptotically anti-de Sitter space, with a
suitable cosmological constant \cite{HP}.
For a black hole, with electric charge $Q$, we replace $V$ with $Q$ and
the analogous condition for the heat capacities\footnote{Strictly speaking it is the enthalpy of the black hole that is the relevant thermodynamic potential when a fixed negative cosmological is present, rather than the
internal energy \cite{KRT}.  At fixed pressure the enthalpy should be a convex function of $S$ and $Q$.}
is then $C_\Phi > C_Q$, where $\Phi$ is the electrostatic potential of the black hole.  In this case the isothermal electrical capacitance, ${\cal C}_T = \left.\frac{\partial Q}{\partial \Phi}\right|_T$, and the isentropic electrical capacitance, ${\cal C}_T = \left.\frac{\partial Q}{\partial \Phi}\right|_S$, are related to the heat capacities by $C_\Phi {\cal C}_S=C_Q {\cal C}_T$, implying ${\cal C}_T > {\cal C}_S$.
For rotating black holes in asymptotically anti-de Sitter space time 
$V$ is replaced with the angular momentum, $J$,
and $-P$ by the appropriate
angular velocity, $\Omega$. The analogue of the compressibilities for a gas
are the moments of inertia, ${\cal I}_S=\left.\frac{\partial J}{\partial \Omega}\right|_S$ and ${\cal I}_T=\left.\frac{\partial J}{\partial \Omega}\right|_T$. For stability $C_\Omega > C_J$ and $C_\Omega {\cal I}_S = C_J {\cal I}_T$ implies ${\cal I}_T >{\cal I}_S$.
The analytic structure of the response functions, in particular the interplay between their zeros and singularities, is crucial in satisfying these relations
and we shall explore this in detail for charged and rotating black holes
in space time dimensions $D\ge 4$. For example the phenomenon of ultra-spinning
black holes in $D>4$ is associated with a negative moment of inertia.

Thermodynamic potentials can also be used to generate a metric on the space of thermodynamic states so that the intrinsic curvature associated with
the metric encodes information about the underlying physics of the thermodynamic system,
\cite{Weinhold,Ruppeiner} (for a review see \cite{RuppeinerReview}).
The study of the geometry of thermodynamic state space for black holes was 
pioneered in \cite{FKG} and has subsequently been investigated by a number of
authors (a non-exhaustive list is \cite{BTZ}-\cite{ML} and there is a review in \cite{RuppeinerBH}). In the present work this programme is taken further and curvatures
are studied for charged and for rotating black holes in $D$-dimensional space-times for all $D\ge 4$, in the presence of a cosmological constant $\Lambda$. 
Generically one expects the intrinsic curvature of thermodynamic state space 
to diverge when response functions, such as heat capacities diverge, \cite{RuppeinerReview,Mirza1}, and this expectation is indeed realised, but we shall see that there is a twist for black holes in that critical points, which are known to exist for $\Lambda<0$, are not visible in the thermodynamic curvatures calculated here.   

When the cosmological constant is included in the set of thermodynamic variables, as argued in \cite{KRT} it should be, the dimension of the thermodynamic state
space is increased.\footnote{The possibility of varying $\Lambda$ was first considered by Henneaux and Teitelboim \cite{HT} and has been re-visited by various authors since \cite{Sekiwa}-\cite{Paddy}. Although the focus here is on
negative $\Lambda$ many of the formulae can be analytically continued to positive $\Lambda$.} Most of the literature to date on the thermodynamic
geometry of black holes has assumed fixed $\Lambda$, but when state space is extended to include $\Lambda$ the dimension of the thermodynamic space is increased and the calculations become more involved.
The thermodynamic geometry of black holes 
with varying $\Lambda$ was considered in \cite{LarranagaMojica}-\cite{ZhangCaiWu2} but the 
study in the present paper is rather different in that
it also considers the role of the thermodynamic volume, the Legendre transform of the pressure, which to our knowledge has not so far been considered as a state variable in the thermodynamic geometry of black holes.

The layout of the paper is as follows. In \S\ref{sec:metrics} the geometry of thermodynamic state space is briefly reviewed;
\S\ref{sec:AdSRN} concerns charged, non-rotating black holes in $D$ space-time dimensions
and extends known results in $D=4$ for spherical black holes
to $D>4$ and general event horizon topologies; \S\ref{sec:MP} discusses the thermodynamic geometry of neutral singly spinning black holes in $D$-dimensions, 
the curvature scalar associated
with the thermodynamic geometry is discussed and singularities in the curvature
scalar are related to singularities in the heat capacity and vanishing compressibility. Finally \S\ref{sec:Conclusions} summarises the conclusions while some
technical results are given in two appendices.

\section{Thermodynamic metrics \label{sec:metrics}}

Different thermodynamic potentials give rise to different thermodynamic metrics
and both the internal energy, $U(S,V)$, and the entropy, $S(U,V)$, can
be used to generate Euclidean signature metrics for thermodynamically stable systems \cite{RuppeinerReview}.  
For example the internal energy of a single component gas
is a convex function of the extensive variables, entropy $S$ and volume $V$,
and the Hessian matrix
\begin{equation}
g_{AB}=\frac{\partial^2 U}{\partial X^A \partial X^B}, \label{eq:WeinholdSV}
\end{equation}
with $(X^1,X^2)=(S,V)$ and $A,B=1,2$, is a positive definite matrix.
Viewed as a metric, originally considered by Weinhold \cite{Weinhold}, 
there is an intrinsic curvature associated with this matrix
which has been calculated for a number of thermodynamic systems
\cite{RuppeinerReview}.

The components of $g_{AB}$ are 
related to the response functions of the thermodynamic system:
\begin{equation}
g_{SS}= \left.\frac{\partial T}{\partial S}\right|_V = \frac{T}{C_V},
\label{eq:CV}
\end{equation}
with $C_V$ the heat capacity at constant volume;
\[
g_{VV}= -\left.\frac{\partial P}{\partial V}\right|_S = \frac{1}{V \kappa_S},
\]
where $\kappa_S=-\frac{1}{V}\left.\frac{\partial V}{\partial P}\right|_S$
is the adiabatic compressibility and
\[g_{SV}=\frac{1}{V\alpha_S}\]
where $\alpha_S=\frac{1}{V}\left.\frac{\partial V}{\partial T}\right|_S$ is the
adiabatic thermal expansion coefficient. Standard thermodynamic relations
can be used\footnote{In particular the aforementioned relation
${C_P \kappa_S}={C_V \kappa_T}$ together with $C_P=C_V +\frac{T V \alpha_P^2}{\kappa_T}$
and $\kappa_T=\kappa_S +\frac{T V \alpha_P^2}{C_P}$,
where  $\kappa_T=-\frac{1}{V}\left.\frac{\partial V}{\partial P}\right|_T$ is the isothermal compressibility, and $\alpha_P= \frac{1}{V}\left.\frac{\partial V}{\partial T}\right|_P$
is the isobaric thermal expansion coefficient.
Also the Maxwell relation 
$\left.\frac{\partial T}{\partial V}\right|_S = -\left.\frac{\partial P}{\partial S}\right|_V$ implies 
\[g_{SV}=-\left.\frac{\partial P}{\partial T}\right|_V\left.\frac{\partial T}{\partial S}\right|_V=
\left.\left(\frac{\partial V}{\partial T}\right|_P 
\left.\frac{\partial P}{\partial V}\right|_T  \right)\left.\frac{\partial T}{\partial S}\right|_V=
 -\frac{\alpha_P}{\kappa_T}\frac{T}{C_V}.\]} 
to show that
\begin{equation}
\det g = \frac{T}{V C_P \kappa_S}.\label{eq:detg}
\end{equation}
A generalisation of this formula when there are more than two independent
thermodynamic variables was given in \cite{Mirza1}.

When response functions diverge the determinant of the metric vanishes
and it will not be invertible.
In a thermodynamic system with two variables the locus of points on
which a response function diverges is called a spinodal curve and
it is natural to ask whether this lack of invertibility on a spinodal curve
is associated
with a genuine singularity in the intrinsic curvature or if it is just a co-ordinate singularity.
Indeed the thermodynamic curvature is inversely proportional $(\det g)^2$ and is expected diverge on spinodal curves \cite{RuppeinerReview,Mirza1}. 
For the van der Waals gas for example the curvature
is proportional to $(C_P\kappa_S)^2$ and diverges when $C_P$ diverges. 
The spinodal curve has two branches in the $P-V$ plane and, with the exception of the critical point where the two branches meet, it lies in a regime that is not thermodynamically stable.  

A related metric, the Ruppeiner metric, is associated with the Hessian matrix of the entropy, $S(U,V)$, which is a concave function, so
\begin{equation} \tilde g_{AB}=-\frac {\partial^2 S} {\partial {\widetilde X}^A \partial {\widetilde X}^B}, \label{eq:RuppeinerUV}\end{equation}
with $({\widetilde X}^1,{\widetilde X}^2)=(U,V)$, is also a positive definite matrix if the system is stable.
It has been argued \cite{RuppeinerReview} that 
the intrinsic thermodynamic curvature
contains information about microscopic physics of the system, for example it diverges when 
response functions diverge. Indeed both curvatures vanish for an ideal gas and both are
positive for a van der Waals gas and diverge at the critical point.    

The Ruppeiner and the Weinhold metrics are con formally related,
\[ \tilde g_{AB}= \frac 1 T g_{AB} \]
and give equivalent information. 
The Ricci scalar of the Weinhold metric, $R$, and that of the
Ruppeiner metric $\widetilde R$, are therefore related by the standard formula for con formally related metrics,
\begin{equation}
\widetilde R = T R + (n-1)\nabla^2 T - \frac{(n-1)(n+2)}{4}\frac{(\nabla T)^2}{T}\label{eq:RRtilde}
\end{equation}
where $n$ is the number of independent thermodynamic variables,\footnote{When the variables are $(S,V)$ then $n=2$, but we quote the more general result.} one of which is $T$. 
The second term on the right-hand side involves 
\begin{eqnarray*}
\nabla^2 T &=& \frac{1}{\sqrt {\det g}} \partial_a\bigl(\sqrt {\det g}\, g^{ab}\partial_b T\bigr)= \frac{1}{\sqrt {\det g}} \partial_a\bigl(\sqrt {\det g}\, g^{ab}g_{bS}\bigr)\\
&=&\frac{1}{\sqrt {\det g}} \left.\frac{\partial \sqrt {\det g}}{\partial S}\right|_V.
\end{eqnarray*}  
The last term on the right-hand side of (\ref{eq:RRtilde}) is related to the heat capacity at constant volume, $C_V=\frac{T}{g_{SS}}$,
\begin{equation}
\frac{(\nabla T)^2}{T}= \frac{1}{C_V},
\end{equation}
since 
\[\nabla_a T = \partial_a T = g_{aS} \qquad  \Rightarrow \qquad 
(\nabla T)^2 = g^{ab}g_{aS}g_{bS}=g_{SS}=\frac{T}{C_V}\] from (\ref{eq:CV}).
When $n=2$ and the only variables are $(S,V)$ the relation between the Weinhold
and the Ruppeiner curvatures is
\begin{equation}
\widetilde R = T R  
- \frac{1}{2}\partial_S\ln \left(\frac{T C_V\kappa_T}{V}\right) \label{eq:RRRW}
\end{equation}
(the derivative here is at constant $V$ and the factor of $V$ has been chosen
to make the argument of the logarithm dimensionless).

Another class of conformally related metric
was proposed in \cite{Quevedo1}, but the focus here will be restricted to 
the specific cases (\ref{eq:WeinholdSV})
and (\ref{eq:RuppeinerUV}), otherwise the analysis would become rather too 
long.


The thermodynamic potentials for a gas also depend on the number of particles
so, for a single component gas consisting of $N$ particles, $U(S,V,N)$, and 
a complete description requires three independent variables, $X^A=(S,V,N)$ with $A=1,2,3$.
One might expect 
that a full description 
of the thermodynamic geometry would require calculating a $3\times 3$ Ricci tensor, but this is not the case in standard thermodynamics as the resulting
$3\times 3$ Hessian matrix is not invertible
and cannot be used to determine a curvature.  
This is a consequence of the Gibbs-Duhem relation, which follows
from homogeneous scaling of the extensive thermodynamic state variables and
potentials.
Consider the Gibbs
free energy,
\begin{equation} {\cal G}(T,P,N) = \mu N = U(S,V,N) - TS + PV,
\label{eq:Gibbs-Duhem}\end{equation}
where
$\mu=\left.\frac{\partial U}{\partial N}\right|_{S,V}$
is the chemical potential. In the 3-dimensional state space parameterised by
extensive variables $X^A=(S,V,N)$ this can be written as
\begin{equation} U(X)=X^A Y_A(X) \label{eq:UxX}
\end{equation}
where $Y_A(X):=\partial_A U$.
Now the $3\times 3$ Hessian matrix is
\[ g_{AB}=\partial_A Y_B\]
so
\[ Y_A= \partial_A U = Y_A + g_{AB} X^B,\]
where (\ref{eq:UxX}) has been used for the second equality.
Hence $g_{AB}X^B = 0$ and the vector $\vec X=(X^1,X^2,X^3)$ is an eigenvector
of $g_{AB}$ with eigenvalue zero (this is essentially a consequence
of homogeneous scaling). Assuming the other two eigenvalues are finite
the $3\times 3$ Hessian matrix is never invertible and is not a suitable
candidate for a for a positive definite metric.
Similarly for a multi-component gas, with $k$ different chemical species $i=1,\ldots,k$ and particle numbers $N_i$ there is always at least one zero eigenvector of the $(k+2)\times(k+2)$ Hessian matrix.
For this reason it is natural to fix the total particle number $N=N_1+\cdots N_k$ and only consider a $(k+1)\times (k+1)$ metric.

For black holes we do not yet have an analogue of particle number,
nevertheless scaling arguments
can be applied to derive the black hole analogue
of the Gibbs-Duhem relation, the Smarr relation \cite{Smarr}.  
For black holes the extensive thermodynamic variables do not scale homogeneously. For example, in $D$ space-time dimensions the black hole mass, $M$, and electric charge, $Q$, have dimension $(D-3)$, (in units with $G=c=1$), while entropy, $S$, and angular momentum, $J$, 
have dimension $(D-2)$ and $\Lambda$ has dimension $-2$.  If $\Lambda$ is equated to a pressure, 
\[ \Lambda = - 8\pi P,\]
so negative $\Lambda$ gives a positive pressure,
then the black hole mass has the thermodynamic interpretation of 
enthalpy \cite{KRT}. 

The mass is a  function of entropy, angular momentum  $J^i$
(where $i=1,\ldots,r$ and $r$ is the rank of $SO(D-1)$),
electric charge and pressure: 
$M(S,J^i,Q,P)$. The thermodynamically conjugate variables are temperature, angular 
velocity $\Omega_i$, electric potential $\Phi$ and thermodynamic volume $V$:
\[ T=\frac{\partial M}{\partial S}, \qquad
\Omega_i=\frac{\partial M}{\partial J^i}, \qquad
\Phi=\frac{\partial M}{\partial Q} \qquad \hbox{and} \qquad
V=\frac{\partial M}{\partial P}.\]
The Smarr relation, the black hole analogue of (\ref{eq:Gibbs-Duhem}), then reads
\begin{equation}
(D-3)M= (D-2)\,\mathbf{\Omega. J}+(D-2)TS  -2PV+(D-3) Q\Phi.
\label{eq:Smarr}
\end{equation}

When $\Lambda$ is allowed to vary
the Weinhold metric should be defined in terms of the internal energy $U$
in order to have a positive definite metric.
The internal energy is the Legendre transform of the enthalpy with respect to $P$,
\[ U(S,J^i,Q,V) = M - PV.\]
Then 
\[ T=\frac{\partial U}{\partial S}, \qquad
\Omega_i=\frac{\partial U}{\partial J^i}, \qquad
\Phi=\frac{\partial U}{\partial Q}, \qquad
P=-\frac{\partial U}{\partial V}\]
and
\begin{equation}
g_{AB}=\frac{\partial^2 U}{\partial X^A \partial X^B},\label{eq:Weinhold}\end{equation}
with $X^A=(S,J^i,Q,V)$ and $A=1,\ldots,r+3$.

One immediate consequence of the inhomogeneous scaling of black hole
thermodynamic functions is that, unlike ordinary thermodynamics, $g_{AB}$
need not have a zero eigenvalue.  However, if $P=0$, it always has a negative
eigenvalue, \cite{Stability}, and there are no thermodynamically stable asymptotically flat black holes in any dimension, regardless of rotation or charge.
On the other hand asymptotically AdS black holes can be stabilized by a large enough positive pressure: Hawking and Page realized this for non-rotating neutral 
black holes \cite{HP} and it was generalized to the charged non-rotating case 
in \cite{Diasetal} and the charged rotating case in \cite{Stability}.

Note that the complete Legendre transform of $U$, 
\begin{equation} \E(\Omega,T,P,\Phi):=U -{\mathbf \Omega}_i {\mathbf J}^i - TS + PV -\Phi Q,\end{equation}
is non-zero for black holes --- in contrast to ordinary thermodynamics
where (\ref{eq:Gibbs-Duhem}), and its generalisation to multi-component systems,
 implies that 
\[ U - TS + PV -\mu_i N^i=0.\]
This is again a consequence of the inhomogeneous scaling of black hole thermodynamic variables and $\E$ is in fact related to the Euclidean action, $I_E$, of the black hole via \cite{Gibbons}
\[\E = T I_E.\]

When discussing the thermodynamic geometry of asymptotically AdS
black holes it is very common in the literature to use the enthalpy 
$M(S,J^i,Q,P)$ rather than the internal energy $U(S,J^i,Q,V)$ to derive a Weinhold-like metric. This not unreasonable when $P$ is fixed and indeed 
for asymptotically flat black holes with $P=0$ it makes no difference which one uses as they are the same. When $P\ne 0$ the $(r+2)$-dimensional metric obtained
from the Hessian of $M$ by varying $S$, $Q$ and $J^i$ with $P$ fixed will be
positive definite for a thermodynamically stable black hole.
Typically there are curvature singularities when the heat capacity diverges. 

In \cite{LarranagaMojica} $P$ was varied along with the other parameters
but this has the disadvantage that the Hessian matrix of $M(S,J^i,Q,P)$
cannot be expected to be positive definite for thermodynamically stable systems,
one should use $U(S,J^i,Q,V)$ to get a positive definite Weinhold metric.  

A technical point is that, for a given solution of Einstein's equations, an explicit expression for the mass in terms
of thermodynamic variables is often cumbersome at best and 
intractable at worst.  It is usually more convenient to express the thermodynamic potentials in terms of variables
other than $(S,J^i,Q,V)$. Curvature tensors can of course
be calculated in any co-ordinate system, invoking general co-ordinate
invariance, but one must be careful because the right hand side of (\ref{eq:Weinhold}) is not co-variant under general co-ordinate transformations: $X^A=(S,J^i,Q,V)$ are a 
privileged set of co-ordinates that picks out a particular Lagrangian sub-manifold of thermodynamic \lq\lq phase space'', \cite{Quevedo1}.  
Writing the thermodynamic line element as
\begin{equation} d^2 s = \frac{\partial ^2 U}{\partial X^A \partial X^B}\,d X^A dX^B
= d Y_A dX^A,
\label{eq:Weinhold-line-element}\end{equation}
where $Y_A=(T,\Omega_i,\Phi,-P)$,
we can change variables from $X^A$ to $X^{A'}(X)$ in terms of which equation
(\ref{eq:Weinhold-line-element}) becomes
\[ d^2 s = g_{A' B'}\, d X^{A'} d X^{B'}, \]
with  
\begin{equation} g_{A' B'}= \frac{\partial Y_C}{\partial X^{A'}}  
\frac{\partial X^C}{\partial X^{B'}}.
\label{eq:Weinhold-general}\end{equation}
The metric thus factorises into a product of two matrices. Equation
(\ref{eq:Weinhold-general}) will prove to be the most practical starting
point for calculating curvatures in an arbitrary co-ordinate system $X^{A'}$
for the thermodynamic state space
when the functional forms of $Y_A(X')$ and $X^A(X')$ are known explicitly.

\section{Asymptotically AdS  Reissner-Nordstr\"om \\
 space-times \label{sec:AdSRN}} 

In this section we consider the thermodynamic geometry of charged, non-rotating
AdS black holes. This was investigated in 4-dimensional space times 
by the authors of  \cite{4D-RN-K-KN-metric}
and here the analysis is extended higher dimensional space-times with general dimension $D$.

With no extra work, and with a view to applications in the AdS/CFT correspondence, we can also incorporate more general event horizon topologies by denoting
the curvature of the event horizon by $k$: so $k=1$ is a spherical event horizon, $k=0$ a flat one (which we take to toroidal for convenience) 
and $k=-1$ the $(D-2)$-dimensional space of constant negative curvature. 

The extensive variables are the electric charge $Q$ and the entropy $S$.
The pressure $P$ will be kept fixed in this section, parameterised for notational
convenience by
\[\lambda = \frac{16 \pi P}{(D-1)(D-2)}.\] 
If $r_h$ is the radius of the outer horizon the entropy,
in units with $G=c=\hbar=1$, is
\begin{equation}
S=\frac{\V r_h^{D-2}}{4},
\end{equation}
where $\V$ is a the volume of the event horizon with $r_h$ set to unity.\footnote{For $k=+1$ 
this is the volume of the unit $(D-2)$ sphere, for $k=0$ and $k=-1$ the event horizon can be made to have finite area by suitably identifying points.}
The mass (enthalpy) of the black hole is
\[
M=\frac{(D-2)\V r_h^{D-3} \left( k + \lambda r_h^2 + \frac{Q^2}{r^{2D-6}}\right)}
{16 \pi}.
\]
The Hawking temperature is
\begin{equation}
T=\frac{(D-1)\lambda r_h^2 +(D-3)\left(k-\frac{Q^2}{r_h^{2D-6}} \right)}{4\pi r_h},\label{eq:HTemp}
\end{equation} 
placing an upper bound $Q^2\le \Bigl(\frac{D-1}{D-3}\lambda r_h^2  + k\Bigr)  r_h^{2D-6}$ on $Q^2$ to ensure $T\ge 0$. 

In \cite{4D-RN-K-KN-metric} a Weinhold metric on the 2-dimensional state space parameterised by $S$ and $Q$ using the Hessian of $M(S,Q,\lambda)$, with $\lambda$ fixed, was considered in $D=4$, with $X^A=(S,Q)$, $A=1,2$
\begin{equation}
g_{AB}=\frac{\partial^2 M}{\partial X^A\partial X^B}.
\label{eq:RN_Hess}
\end{equation}
With $\lambda$ fixed this is a positive definite metric for a thermodynamically stable black hole.
It gives rise to a geometry with a positive scalar curvature for $D=4$, 
\cite{4D-RN-K-KN-metric}, and this result generalizes
to $D\ge 4$.  The metric is written here in $(r_h,Q)$ co-ordinates,
so as to avoid fractional powers of $S$: using 
(\ref{eq:Weinhold-general}) it evaluates to
\begin{equation}
g= \frac{(D-2)\V}{8\pi}
\left( \begin {array}{cc} 
 r_h^{D-5}Z_Q(r_h,Q,\lambda)  &-{\frac { \left( D
-3 \right) Q}{r_h^{D-2}}}\\\noalign{\medskip}-{\frac { \left( D-3 \right) Q}{r_h^{D-2}}}&
{\frac {1}{ {r_h}^{D-3}}}
\end {array} \right), \label{eq:RN_W_metric}
\end{equation}
where $Z_Q(r_h,Q,\lambda)$ is 
\begin{equation}
Z_Q(r_h,Q,\lambda)=(D-1)\lambda r_h^2  -2\pi r_h T + (D-3)^2\frac{Q^2}{r_h^{2D-6}}.
\end{equation}

The determinant of $g$ vanishes when the heat capacity diverges, but we need to be careful to distinguish between the heat capacity at constant charge $C_Q$ and the heat capacity at constant electric potential $C_\Phi$. 
In this context the electric potential is the thermodynamic conjugate of the charge,
\[ \Phi=\left.\frac{\partial M}{\partial Q}\right|_S=\frac{(D-2)\V \,Q}{8\pi r_h^{D-3}},\]
and the heat capacity at constant $\Phi$ is 
\begin{equation}
C_\Phi =\frac{(D-2)\V \pi r_h^{D-1} T}
{2 Z_\Phi(r_h,Q,\lambda)}
\label{eq:CPhi}
\end{equation} 
where
\begin{equation}
Z_\Phi(r_h,Q,\lambda)= (D-1)\lambda r_h^2 - 2\pi r_h T.
\end{equation}
Indeed 
\[ \det g = \frac{\V^2 (D-2)^2 Z_\Phi(r_h,Q,\lambda)}{64 \pi^2 r_h^2}\] 
and a necessary condition for stability is
\begin{equation}
 (D-1)\lambda r_h^2> 2\pi r_h T,
\end{equation} 
which is always true for $k\le 0$
and only imposes a genuine constraint for $k=1$.
The  heat capacity at constant charge is
\begin{equation}
C_Q=
\frac{(D-2)\V \pi r_h^{D-1} T}{2 Z_Q(r_h,Q,\lambda)}
\end{equation}
and the analogue of $C_P > C_V$ for a gas is $C_\Phi > C_Q$. Since $Z_Q > Z_\Phi$
this is satisfied provided $Z_\Phi$ is not negative.
Note that both $C_Q$ and $C_\Phi$ are finite for $k\le 0$, the heat capacity can only become singular if $k=1$.  

The adiabatic electrical capacitance,
\[\cap =\left.\frac{\partial Q}{\partial \Phi}\right|_S = \frac{8\pi r_h^{D-3}}{(D-2) \V },
\]
is independent of the electric charge, depending only on the entropy,
and is always positive.
The isothermal electrical capacitance  ${\cal C}_T$ immediately follows from 
$C_Q{\cal C}_T= C_\Phi{\cal C}_S$.

There is a critical point if $\frac{\partial^2 M}{\partial S^2}$ and $\frac{\partial^2 M}{\partial S^2}$ vanish simultaneously. This cannot happen for $k=0$ or $k=-1$, but for $k=+1$ there
is a critical point at
\begin{equation}
r_{h,*}^2=\frac{(D-3)^2}{(D-1)(D-2)\lambda},\qquad \frac{Q_*^2}{r_{h,*}^{2D-6}}=\frac{1}{(D-2)(2D-5)},
\label{eq:Qcrit}
\end{equation}
found in \cite{ChamblinEmparanMyersJohnson} \cite{ChamblinEmparanMyersJohnson2}.
By definition $C_Q$ diverges at the critical point, but $C_\Phi$ is finite and negative there.
Thus $C_\Phi <0$ at the critical point, which is therefore in a thermodynamically
unstable regime unless $Q$ is held fixed.

The intrinsic scalar curvatures arising from the Weinhold and Ruppeiner geometries
are described in detail in appendix \ref{app:RNADS}.  For the Weinhold metric (\ref{eq:RN_W_metric}) the Ricci scalar takes rather a simple form,
\begin{equation} R=  \frac{ (D-3)^2}{(D-2)}\frac{\pi r_h k}{S Z_\Phi^2}.
\label{eq:RNADS_W}
\end{equation}
Thus the curvature of the event horizon is reflected in the sign of $R$,
in particular the Weinhold metric is flat for $k=0$.

The curvature scalar associated with the corresponding Ruppeiner metric, with $\Lambda$ fixed, follows from the charged black hole analogue of (\ref{eq:RRRW})
for a gas,
\begin{equation}
\widetilde R = T R - \frac{1}{2}\left.\frac{\partial}{\partial S}\right|_Q 
\ln\left(\frac{ T C_\Phi {\cal C}_S}{Q^2}\right),
\end{equation}
or by direct calculation from the Ruppeiner metric.
It is proportional to $\lambda$ and therefore vanishes
in asymptotically flat space-time: explicitly
\begin{equation}
\widetilde R=\frac{(D-1)\lambda \bigl\{3\pi T - (D-1) r_h\lambda \bigr\}\widetilde F(r_h,\lambda,Q) C_\Phi^2}
{2\bigl\{2 \pi (D-2) S T\bigr\}^3}
\end{equation}
where 
\[\widetilde F(r_h,\lambda,Q) = 2(D-1)\lambda r_h^2 + 4 (D-4) \pi r_h T +2(D-3)^2 \frac{Q^2}{r_h^{2D-6}}\] 
is a positive function.
That $\widetilde R$ vanishes for asymptotically flat space-time
in $D=4$ was observed in  \cite{4D-RN-K-KN-metric} and we see here that this
statement generalises to
$D>4$. $\widetilde R$ can be of either sign when $\lambda>0$ and is negative for
small temperatures, diverging to minus infinity for extremal black holes.

We see that a flat event horizon gives $R=0$ while
a flat cosmology, $\Lambda=0$, gives $\widetilde R=0$.
It was observed in \cite{RuppeinerBH} that repulsive microscopic forces tend to give negative Ruppeiner curvature\footnote{Our convention for the sign of the Ricci scalar is that a sphere has positive curvature: this is
opposite to that of \cite{RuppeinerBH}.} while attractive forces give positive curvature,  but in the absence of an underlying theory of quantum gravity it is not at all clear whether or not this interpretation accounts for  the changing sign of $\widetilde R$ for Reissner-Nordstr\"om black holes when $\lambda = \frac{3\pi T}{(D-1)r_h}$.  

The dependence of the Weinhold scalar on $k$ in (\ref{eq:RNADS_W})
can also be given a thermodynamic interpretation. The sign of $k$ is significant for stability as there is no Hawking-Page phase transformation for $k\le 0$.
This can be seen by calculating the Gibbs free energy for the black hole,
\[ {\cal G}(T,\Phi)=M(S,Q)-TS-Q\Phi,\]
which, in terms of $\lambda$, $r_h$ and $Q$, is
\[ {\cal G}=\frac{\V r_h^{D-3}}{16\pi} \left(k-\lambda r_h^2 - \frac{Q^2}{r_h^{2D-3}}\right).\]
The black hole is only stable against Hawking-Page decay if its Gibbs free energy is less than 
the Gibbs free energy of anti-de Sitter space filled with pure thermal radiation
at the Hawking temperature of the black hole and, if the back reaction of the radiation is ignored, the latter is zero. Thus the black hole is stable against 
the Hawking-Page phase transition if the Gibbs free energy is negative, \cite{HP},  which is
always the case for $k\le 0$.  A Hawking-Page phase transition is only possible if $R>0$ .

Note the critical point is not reflected in either of the curvature scalars
because the singularities of $R$ and $\widetilde R$ are determined
by $(C_Q {\cal C}_T)^2 = (C_\Phi {\cal C}_S)^2$ and there is a locus of zeros
in the isothermal electrical capacitance ${\cal C}_T$ which exactly cancels the singularities in $C_Q$. The fact that the analytic
structure of the curvatures is determined by $C_\Phi$ and not by $C_Q$ is a phenomenon which also occurs for rotating black holes as we shall see in the next section.  The same phenomenon is also present for the van der Waals gas where the
singularities in the curvature are determined by $C_P$ and not $C_V$, the latter is in fact finite. For the van der Waals gas however the critical point is also determined
by $C_P$ and not $C_V$, while for the black hole the critical point is determined by $C_Q$ and not $C_\Phi$, even though $C_\Phi$ can have singularities when $k=1$. 

If the pressure is included as a thermodynamic variable we should Legendre transform from $M(S,Q,P)$ to $U(S,Q,V)=M+PV$ to ensure a positive definite metric for
thermodynamically stable black holes. However the Weinhold and Ruppeiner metrics associated with the 3-dimensional state space,
parameterised by $(S,Q,V)$, are singular because $S$ and $V$ are not independent for non-rotating black holes \cite{BPDVolume}.  This restriction is removed when the black hole rotates and we study this case in the next section.

\section{Asymptotically AdS Myers-Perry black holes \label{sec:MP}}

\subsection{Singly spinning black holes}

The thermodynamic state space associated with a $D$-dimensional rotating black hole is multi-dimensional in general, because of the increasing number of angular momenta as $D$ increases. 
For simplicity we set
$Q=0$ and focus on the singly spinning case where only one of the $J^i$ is
non-zero, $J^1=J\ne 0$ say, with $J^i=0$ for $i=2,\ldots,r$.
The thermodynamic state space is then 3-dimensional, parameterised by $(S,J,V)$.
The geometry can be parameterised by the cosmological constant, a rotational parameter $a$ and the radius of the event horizon $r_h$. 

Details are given in appendix \ref{app:KADS} and here we summarise the
relations between the geometric parameters and the thermodynamic ones.
It is convenient to define dimensionless variables $\a = \frac{a}{r_h}$ and $\l = \lambda r_h^2$
in terms of which the thermodynamic variables $(S,J,V)$ are, \cite{GLPP} \cite{GLPP2}, 
\bea
S&=& \frac{\V r_h^{D-2}}{4} \frac{(1 + \a^2)}{(1-\l a^2)},\nonumber\\
J&=&\frac{\V r_h^{D-2}}{8\pi}\frac{(1+\a^2)(1+\l) \a}{ (1-\l\a^2)^2},\label{eq:SJV}\\
V&=&  \frac{\V r_h^{D-1}}{(D-1)(D-2)}
\frac{(1+\a^2)\bigl\{D-2 -(D-3)\l \a^2 +\a^2 \bigr\}}{(1-\l\a^2)^2},\nonumber
\eea
where $\V$ is the volume of a unit $(D-2)$-sphere.
The conjugate variables $(T,\Omega,P)$ are
\bea
T&=& \frac{(D-3) +(D-5)\a^2 + \l\bigl\{D-1 + (D-3)\a^2\bigr\}}
{4\pi r_h (1 + \a^2)},\nonumber \\
\Omega&=& \frac{(1+\l)\a}{(1+\a^2)r_h},\label{eq:TOP} \\
P&=&\frac{(D-1)(D-2)\l }{16\pi r_h^2},\nonumber
\eea
and the thermodynamic potentials are
\bea
M&=&\frac {\V r_h^{D-3}}{16 \pi }
\frac{(1+\a^2)(1+\l) \bigl\{D-2 - (D-4)\l\a^2\bigr\}}{(1-\l\a^2)^2},\label{eq:MU}\\
U&=& M-PV = \frac{\V r_h^{D-3}}{16\pi}\frac{(1+\a^2)\bigl\{D-2 -(D-3)\l\a^2 + \l^2\a^2\bigr\}}{(1-\l\a^2)^2}.\nonumber
\eea

All extensive quantities ($S$, $J$, $V$, $M$ and $U$) diverge when
$\l\a^2=1$, which should be viewed as the edge of thermodynamic state space
\cite{HawkingTaylorRobinson}, in particular the entropy becomes negative
for $\l\a^2>1$ only $\l\a^2<1$ makes sense thermodynamically.
 
\subsection{Thermodynamic geometry}

\subsubsection{Fixed pressure}

For $\Lambda=0$ singly spinning black holes the thermodynamic state space is 2-dimensional, the independent variables being either $S$ and $J$, for the Weinhold metric, or $M$ and $J$ for the Ruppeiner metric.
The intrinsic curvatures associated with both the 
Weinhold and the Ruppeiner metrics were calculated in \cite{D-RN-Kerr}
and the Weinhold metric is flat for all $D$.

When $\Lambda<0$ and held fixed the space-time can be parameterised either by the geometric variables $(X^{A'})=(r_h,\a)$
or by the thermodynamic variables $(X^A)=(S,J)$: for calculations involving the
thermodynamic metric the former are more useful because (\ref{eq:SJV}), (\ref{eq:TOP}) and (\ref{eq:MU}) are simple ratios of polynomials of fairly low
order in these variables which makes the algebra more tractable.
With the explicit expressions (\ref{eq:SJV}) and (\ref{eq:TOP})  it is straightforward to 
evaluate the matrices $\frac{\partial X^C}{\partial X^{B'}}$ and 
 $\frac{\partial Y_C}{\partial X^{A'}}$, with $Y_C=(T,\Omega)$, and hence determine the
metric (\ref{eq:Weinhold-general}) and its curvature.

Details of the Weinhold and the Ruppeiner curvature scalars at fixed pressure
are given in appendix \ref{app:KADS}.  
They diverge on the spinodal curve for
the heat capacity at constant angular momentum, which is a
\begin{equation}
C_{\Omega,P}=-\frac{4\pi r_h T S \bigl\{D-2-(D-4)\a^2 \bigr\}}
{Z_\Omega(\lambda,\a)}\label{eq:COmega}
\end{equation}
with
\begin{equation}
Z_\Omega(\l,\a)=(D-3)(1+\l\a^2)  -(D-1)\l -(D-5)\a^2.\label{eq:Spinodal}
\end{equation}
The spinodal curve, $Z_\Omega=0$, is therefore given by
\begin{equation} 
\l(\a)={ \frac{ (D-3)-(D-5)\a^2}{(D-1)-(D-3)\a^2}}:=\l_\infty.
\label{eq:Rinfinite}
\end{equation}
In contrast the heat capacity at constant angular momentum is
\begin{equation}
C_J=-\frac{4\pi r_h T S (1+\a^2)^2 \bigl\{D-2+(D-4)\lambda\a^2 \bigr\}}{Z_J(\l,\a)}
\end{equation}
where $Z_J$ is quadratic in $\lambda$ (the explicit expression is given explicitly in appendix \ref{app:KADS}, equation (\ref{eq:ZJ})).  In the thermodynamically stable regime $C_\Omega>C_J$.

There is a critical point and a second order phase transition
associated with fixed $J$, first found in \cite{CCK} for $D=4$ but which has an analogue in
any $D$. This critical point is not visible for fixed $\Omega$ as $C_\Omega$ is finite 
there and, in parallel with the Reissner-Nordstr\"om case, 
$C_\Omega$ is in fact negative at the critical point. 

Explicitly the Weinhold curvature is
\beq
R=\frac{16\pi \l(1-\l\a^2)F_1(\l,\a)}{\V r_h^{D-3}(1+\a^2) \{D-2 +(D-4)\l\a^2  \}^2 Z_\Omega^2(\l,\a)}\label{eq:R_Weinhold_JP}
\eeq
with $F_1(\l,\a)$ a polynomial linear in $\l$ and quartic in $\a^2$, given explicitly in 
appendix \ref{app:KADS} equation (\ref{eq:F1}), while the Ruppeiner curvature is
\beq
\widetilde R= -\frac{(D-3)(1-\l^2 \a^4)
\widetilde F_1(\l,\a)\widetilde F_2(\l,\a)}
{\pi \V r_h^{D-1}  (1+\a^2)^2 T \bigl\{D-2+(D-4)\a^2 \bigr\} Z_\Omega^2(\l,\a)},\label{Ruppeiner_Scalar_fixed_P}
\eeq
with $\widetilde F_1(\l,\a)$ a linear in $\l$ and $\a^2$,
and $\widetilde F_2(\l,\a)$ quadratic in $\l$ and $\a^2$
(appendix \ref{app:KADS}, equations (\ref{eq:F1_tilde}) and (\ref{eq:F2_tilde})).

The Weinhold metric is 
no longer flat when $\Lambda<0$, for any $D\ge 4$,
and can be of either sign. Both the Weinhold and the Ruppeiner metrics are 
flat on the line $\l\a^2=1$, which is the boundary of the region
where a thermodynamic interpretation of the black hole is consistent 
as the entropy, angular momentum and mass all diverge on this  line.
Both are singular on the spinodal curve for the heat capacity at constant angular velocity (\ref{eq:COmega}).

\begin{figure}[!ht]
\centerline{\includegraphics[width=8cm]{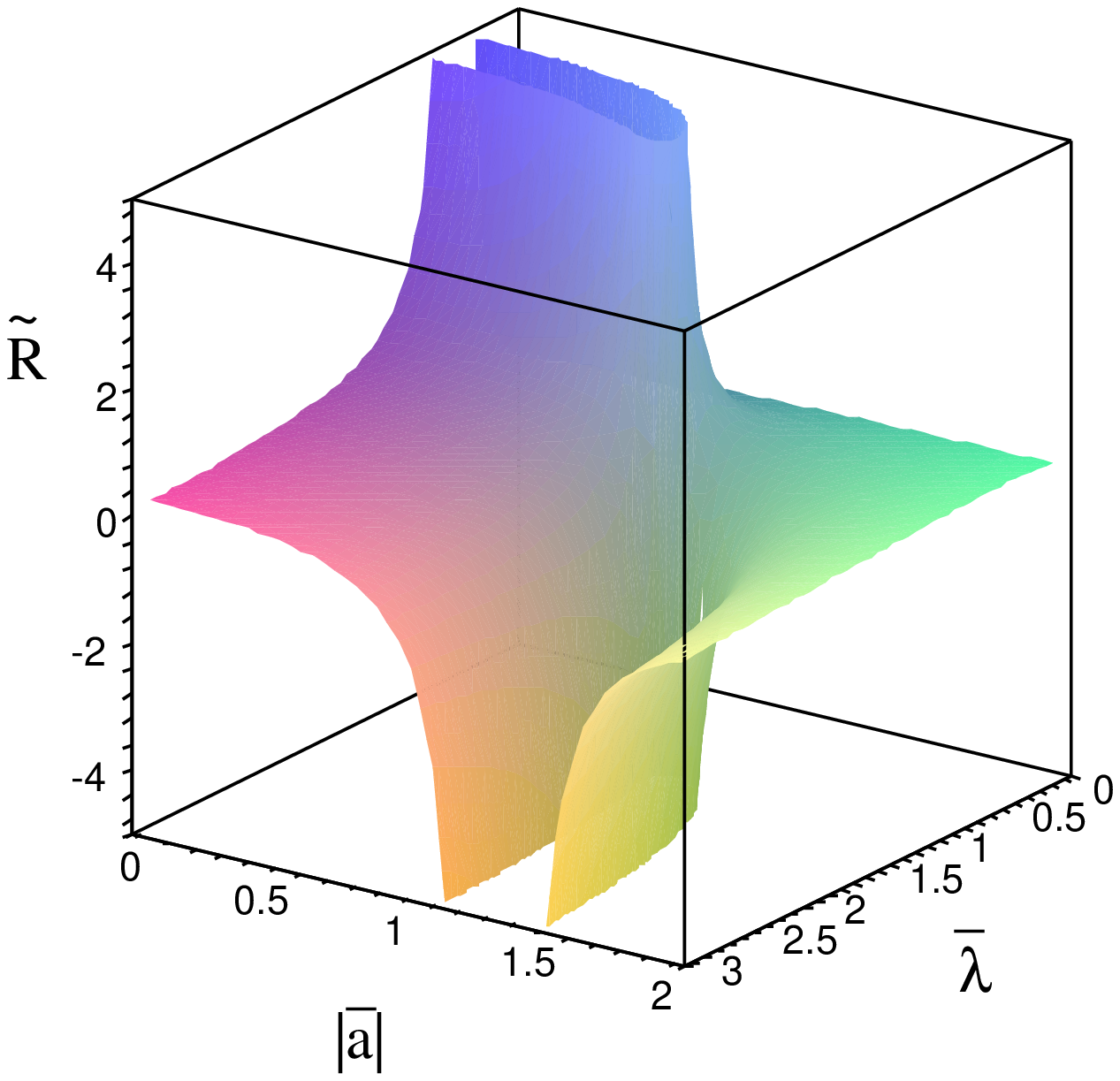}\includegraphics[width=8cm]{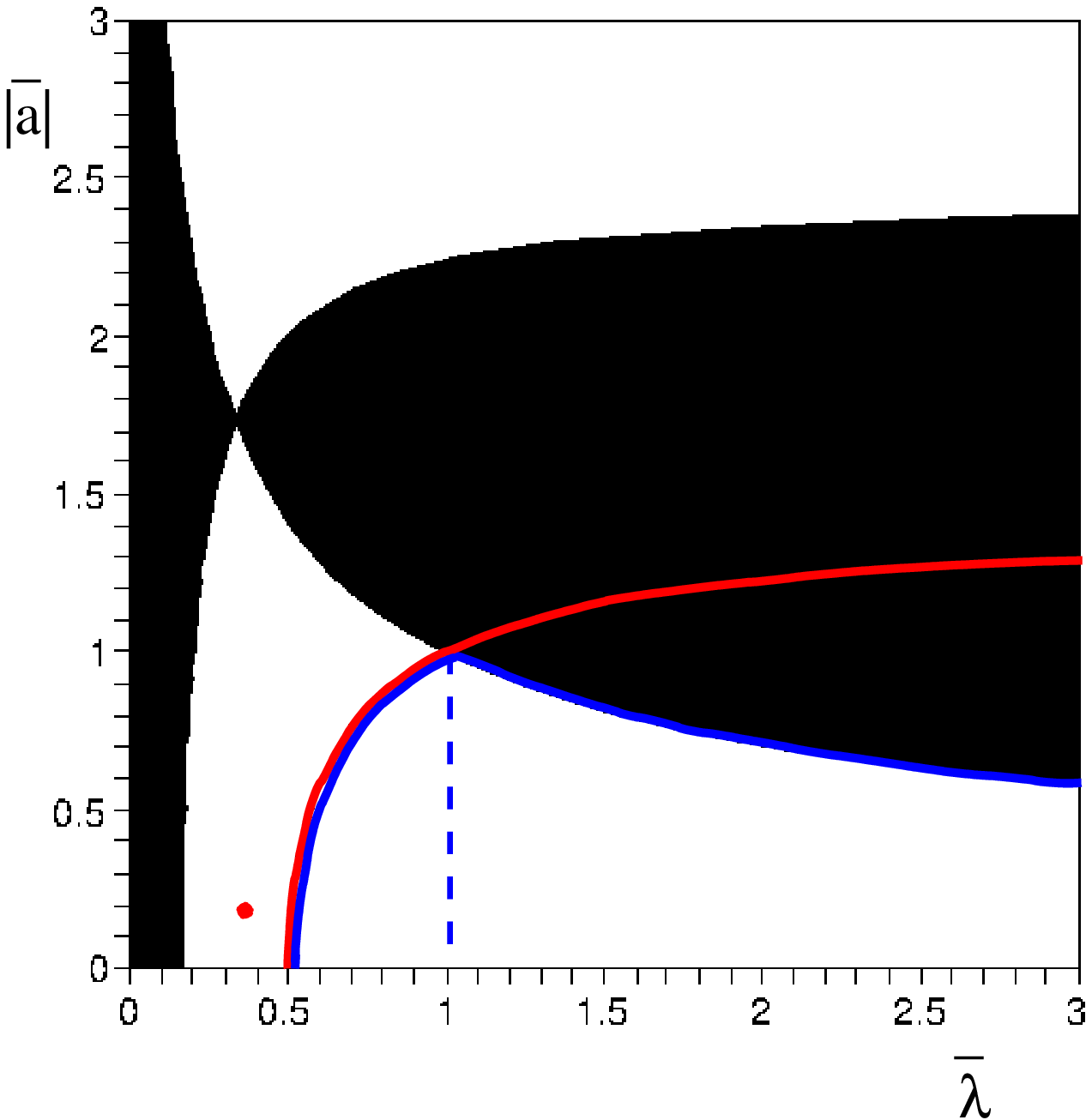}} \caption{The Ricci scalar for the Ruppeiner geometry associated with
a singly spinning asymptotically AdS Myers-Perry black hole in $D=5$
at fixed $\l$, as a function of $\l$ and $\a$
(multiplied by $r_h^3$ to render it dimensionless).
In the right-hand figure $\widetilde R$ is positive in the white regions and 
negative in the black regions. 
The regime of local thermodynamically stability (determined by convexity $M(S,J,P)$ with $P$ fixed) lies below the blue curve,
bounded by the hyperbola $\l\a^2=1$ (blue) and the spinodal curve $Z_\Omega=0$
(red). 
The vertical dashed line indicates the Hawking-Page phase transition ---
the black hole is stable against the Hawking-Page phase transition only
for $\l>1$. The critical point lies in the unstable regime if the angular momentum is not fixed and is indicated by the dot}
\label{fig:RicciRP}
\end{figure}

As an example the Ricci scalar for the Ruppeiner metric is shown in 
Figure \ref{fig:RicciRP}, for $D=5$ (scaled by $r_h^3$ to make it dimensionless). It is negative for small $\l$, for any value of the rotation parameter $\a$,
but is positive everywhere in the thermodynamically stable regime.
The latter observation is also true in $D=4$, \cite{4D-RN-AdS2}\cite{4D-Kerr-AdS2}\cite{4D-Kerr-AdS1}.
  
The singularities in both the Weinhold and the Ruppeiner curvatures are determined by those
of $C_\Omega$ and not $C_J$. In
the context of rotating black holes we have $C_J {\cal I}_T =C_\Omega {\cal I}_S$,
where ${\cal I}_S$ and ${\cal I}_T$ are the isentropic and isothermal
moments of inertia respectively, equations (\ref{eq:I_S}) and (\ref{eq:I_T})
in appendix \ref{app:KADS}.
The curvatures here are proportional to $(C_J {\cal I}_T)^2 = 
(C_\Omega {\cal I}_S)^2$: there are singularities in $C_J$ that
are cancelled by corresponding zeros in the isothermal moment of inertia\footnote{In this section $\lambda$ is held fixed, so it is implicit that the moment of inertia here is
calculated at fixed pressure ${\cal I}_T={\cal I}_{T,P}$.} 
${\cal I}_T$ and singularities in ${\cal I}_T$ which are the same as those in $C_\Omega$.
Note also that the singularities of
\begin{equation}
{\cal I}_S= 
\left.\frac{\partial J}{\partial \Omega}\right|_S=
\frac{r_h S (1+a^2)^2 \bigl\{D-2+(D-4)\lambda\a^2 \bigr\} }{2 \pi (1-\l \a^2)^2\bigl\{ D-2 -(D-4)\a^2\bigr\}},
\end{equation}
in the region with $\l\a^2<1$,
are cancelled by zeros in $C_\Omega$ and the only singularities in the
curvatures are those of $C_\Omega$. 
For $D\ge 5$, $C_\Omega$ is positive if either $Z_\Omega>0$ and $a^2<\frac{D-2}{D-4}$ or $Z_\Omega<0$ and 
$\a^2 > \frac{D-2}{D-4}$, but the latter possibility necessarily implies
that ${\cal I}_S<0$:
for ultra-spinning black holes the moment of inertia is negative.

In summary, when the pressure is held fixed, both the Weinhold and the Ruppeiner
scalar curvatures diverge on the spinodal curve for $C_\Omega$, $Z_\Omega=0$ in (\ref{eq:Spinodal}), and these are the only singularities.
The curvature singularities are determined by those of $C_\Omega$, not $C_J$.  

\subsubsection{Varying pressure}

Allowing for varying $\Lambda$ there is potentially a 4-dimensional space of thermodynamic states associated with singly spinning charged black holes, parameterised by $(S,Q,J,V)$.  A completely general analysis would be rather complicated and we restrict ourselves here to studying the 3-dimensional state space of singly spinning electrically neutral black holes. Even then, with a 3-dimensional state space, the full Ricci tensor is necessary for
a complete description of the curvature and for simplicity the discussion here
is restricted to 
to the Ricci scalar, for which the properties are not too difficult to extract.

The diagonal components of the Weinhold metric in thermodynamic co-ordinates
$(S,J,V)$ have direct physical interpretations:
\bea
g_{SS}&=&\frac{\partial T}{\partial S}=\frac{T}{C_{V,J}}\nonumber \\
g_{JJ}&=&\frac{\partial \Omega}{\partial J}={\cal I}_{S,V}^{-1},\nonumber\\
g_{VV}&=&-\frac{\partial P}{\partial V}=\frac{1}{V\kappa_{S,J}}\nonumber 
\eea
where ${\cal I}_{S,V}$ is the isentropic moment of inertia of the black hole, with $V$ fixed. 

The Weinhold and Ruppeiner metrics of a singly spinning asymptotically AdS Myers-Perry black hole, with internal energy 
$U(S,J,V)$,
can be evaluated in $(r_h,\a,\l)$ co-ordinates using (\ref{eq:Weinhold-general}).  They are found to be
\begin{equation} R=\frac{16\pi}{(D-3) \V r_h^{D-3}} \frac{(1-\l \a^2)F_2(\l,\a^2)}{\a^4 ( 1+{\a}^{2})\bigl( 1+\l\bigr)^{2} Z_\Omega^2(\l,\a),}\end{equation}
where $F_2$ is quadratic in $\l$ and quartic in $\a^2$
(more details are given in appendix \ref{app:KADS}). 

\begin{equation}
\widetilde R = \frac{(1-\l\a^2)\widetilde F_3(\l,\a)}
{4\pi (D-3)\V r_h^{D-1} \a^4 (1+\a^2)^3 (1+\l)^2 T  Z_\Omega^2},
\end{equation}
where $\widetilde F_3(\l,a)$ is a quintic polynomial in $\l$ and sixth order in $\a^2$.

Both curvatures diverge for
$\a^2=0$ and on the curve $\l_\infty$. The former singularity, at $\a^2=0$, is associated with the incompressibility of a non-rotating black hole \cite{Compressibility} while the latter is again the same locus of points
as that on which the heat
capacity at constant angular velocity and pressure $C_{\Omega,P}$ diverges \cite{Stability}.

\begin{figure}[!ht]
\centerline{\includegraphics[width=8cm]{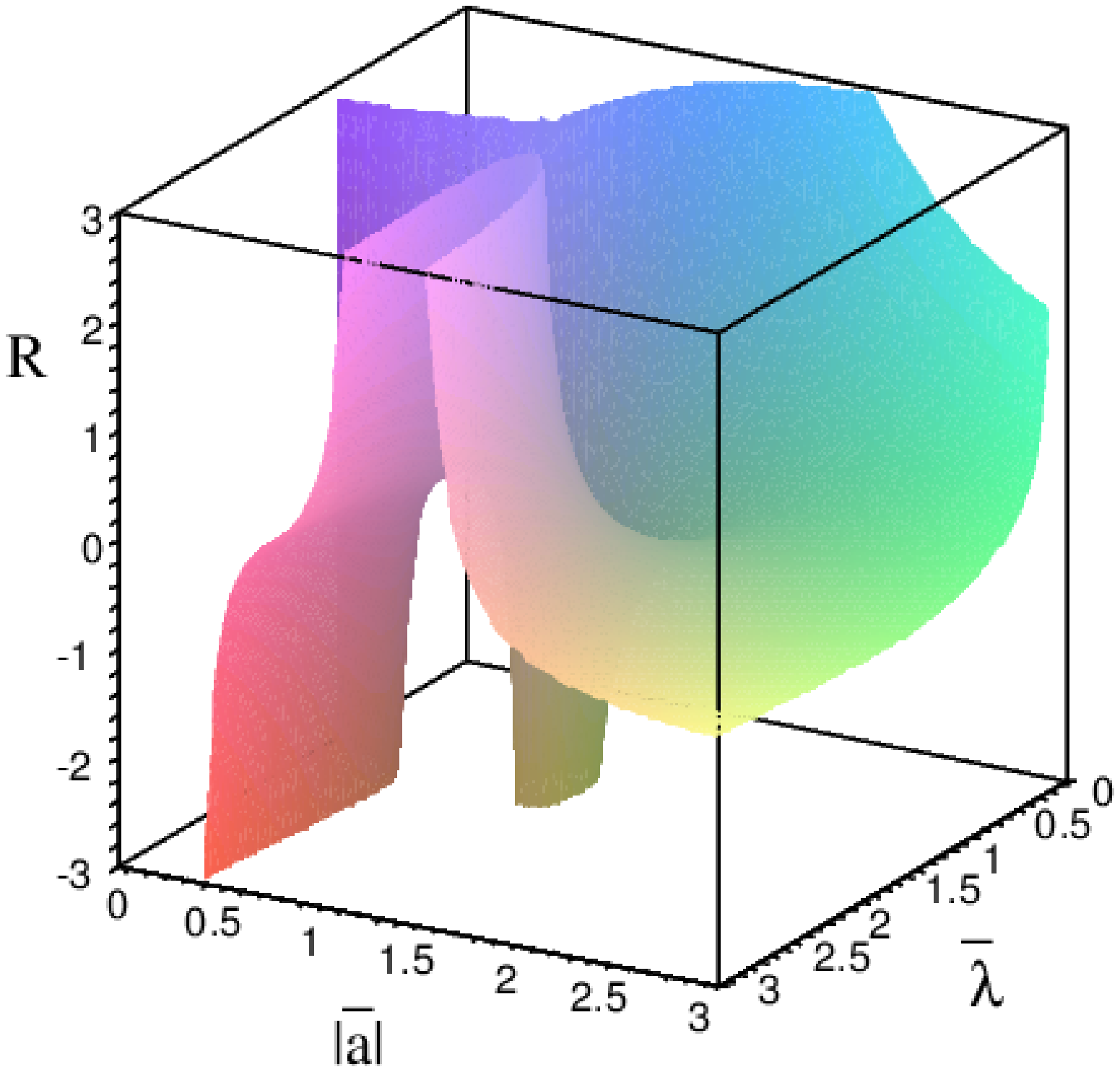}\includegraphics[width=8cm]{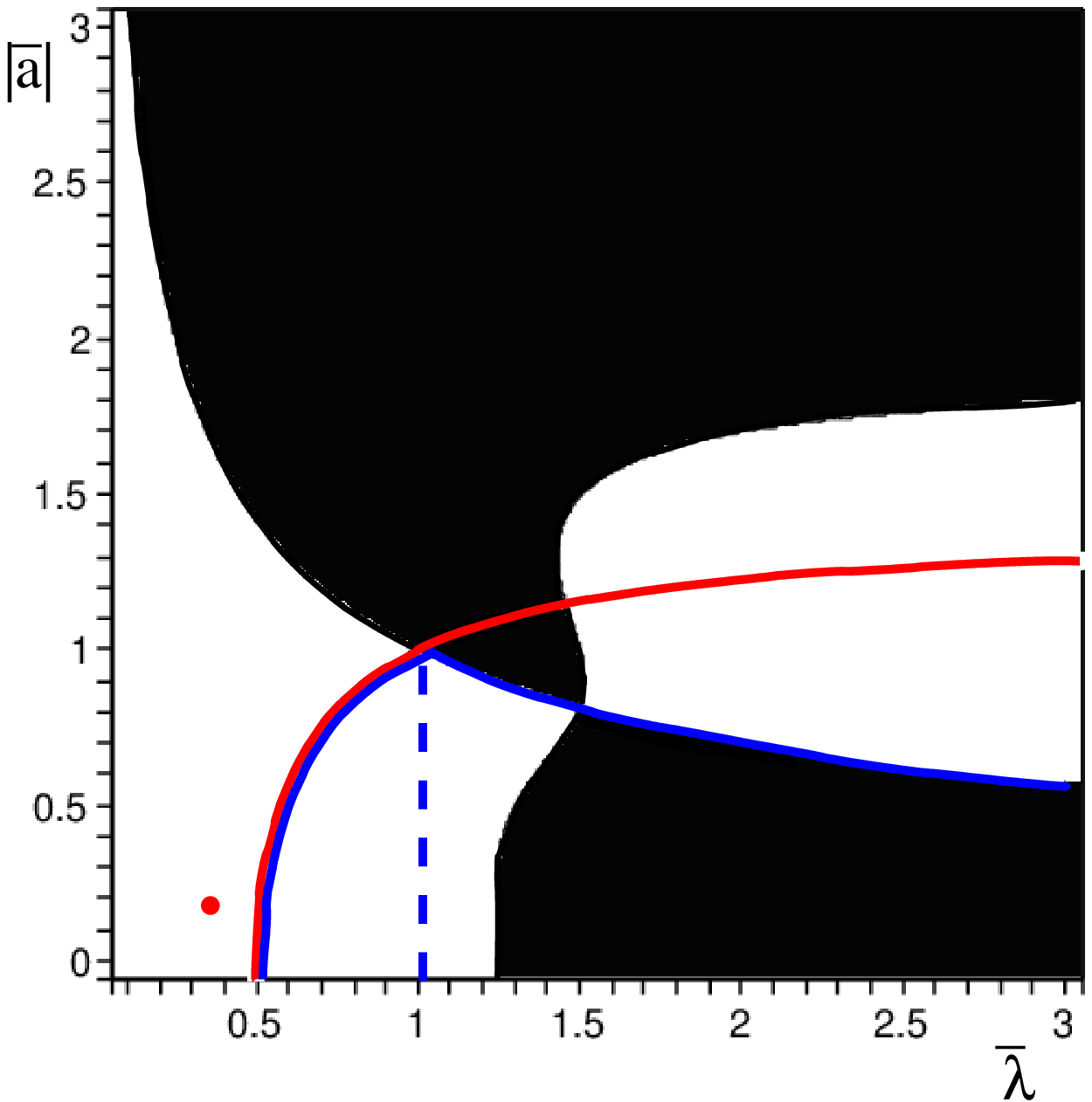}} \caption{The Ricci scalar for the Weinhold geometry associated with
a singly spinning asymptotically AdS Myers-Perry black hole in $D=5$, as a function of $\l$ and $\a$ (multiplied by $r_h^2$ to render it dimensionless).
The colour coding is the same as in \ref{fig:RicciRP}.}
\label{fig:RicciW}
\end{figure}

\begin{figure}[!ht]
\centerline{\includegraphics[width=8cm]{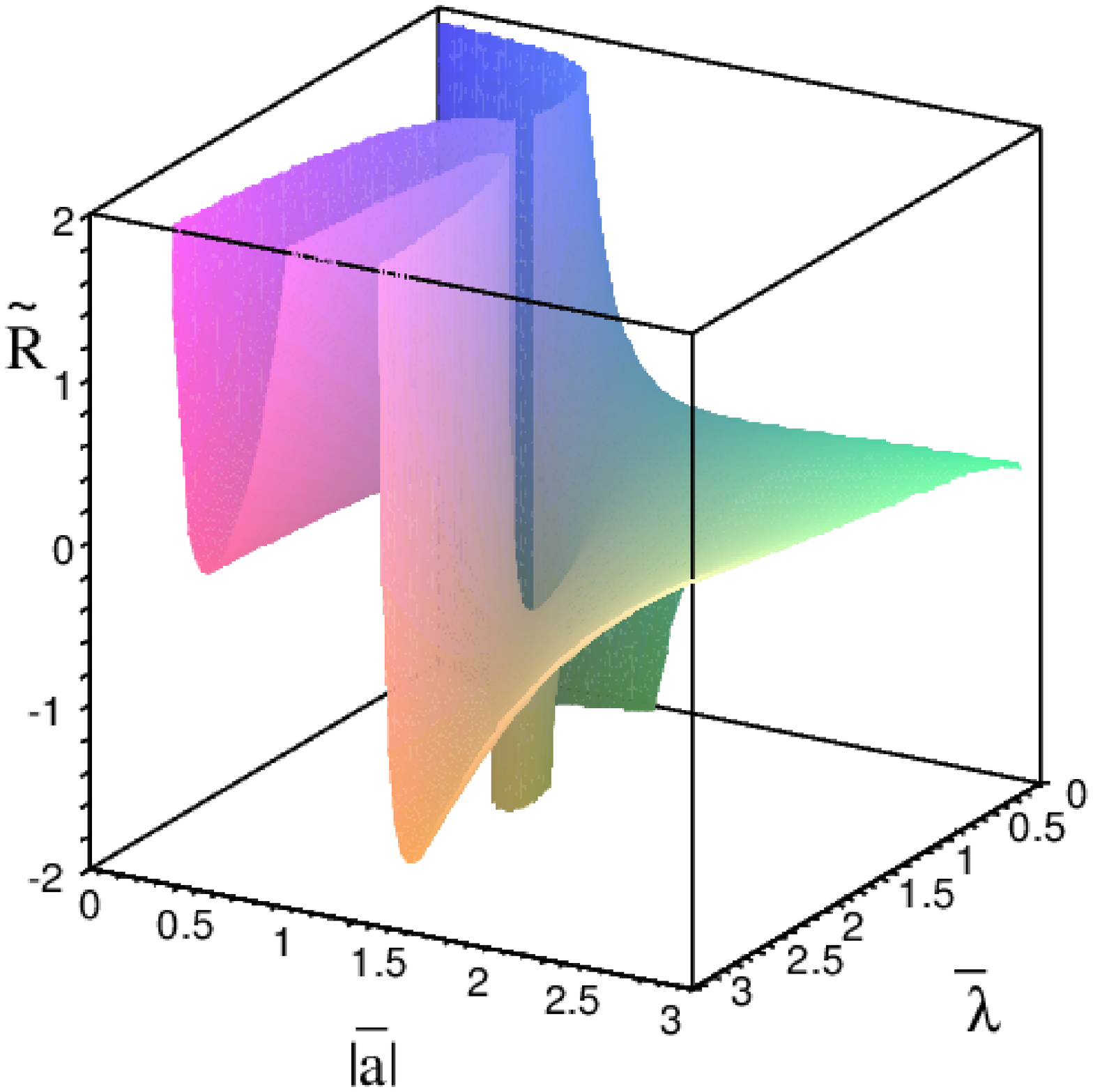}\includegraphics[width=8cm]{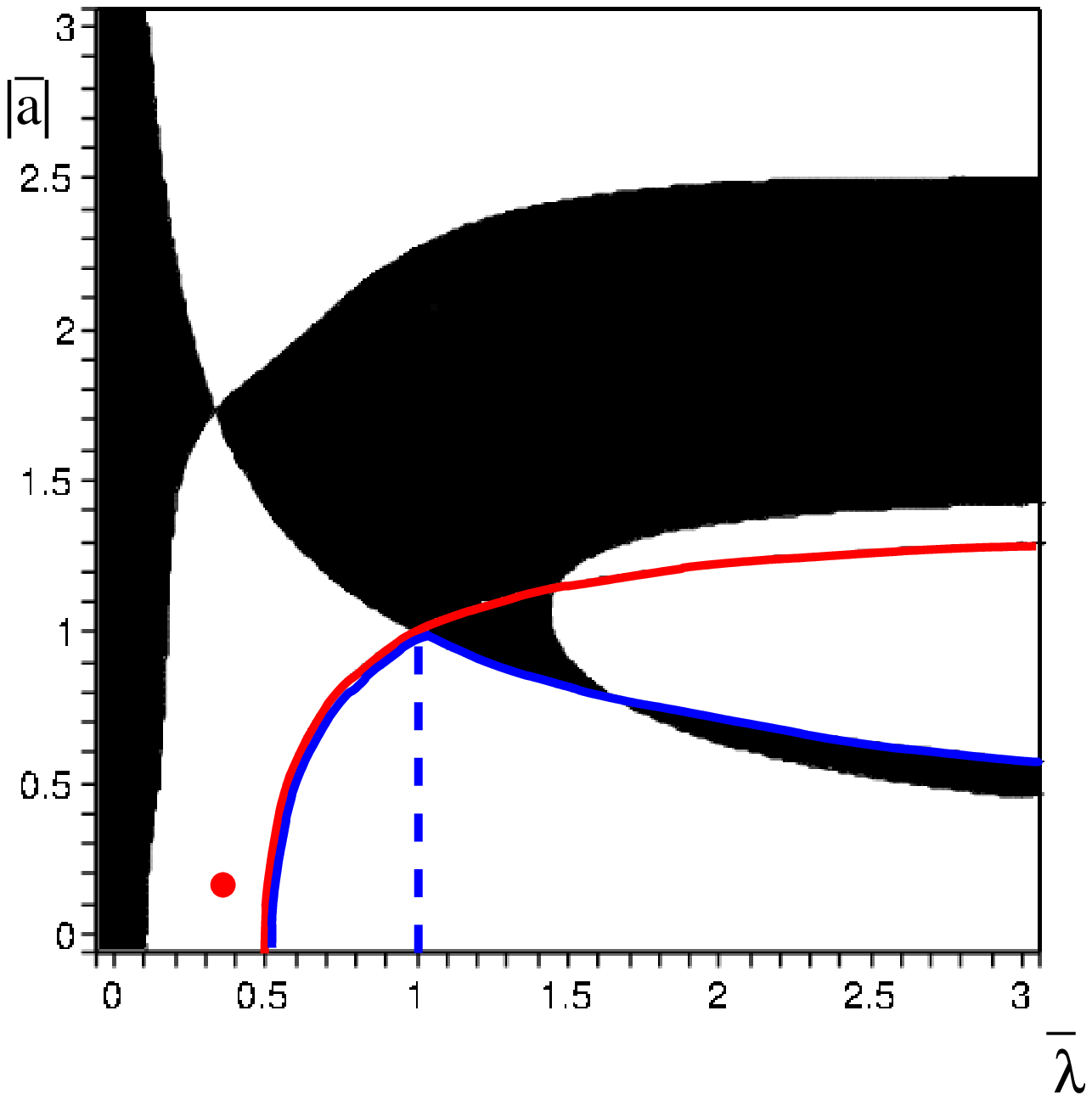}}
\caption{The Ricci scalar for the Ruppeiner geometry associated with
a singly spinning asymptotically AdS Myers-Perry black hole in $D=5$, as a function of $\l$ and $\a$
(multiplied by $r_h^3$ to render it dimensionless). The colour coding is the same as in Figure \ref{fig:RicciRP}.} 
\label{fig:RicciR}
\end{figure}

For illustrative purposes we show the Weinhold Ricci scalar, $R$, for $D=5$ 
in Figure \ref{fig:RicciW}, scaled by a factor $r_h^2$ to make it a dimensionless function
of only two variables $\a$ and $\l$. 
For comparison the Ruppeiner Ricci scalar
$\widetilde R$ is shown in Figure \ref{fig:RicciR}, scaled by $r_h^3$. 
In both cases the Ricci scalar diverges on the curve $\l_\infty$ in equation
(\ref{eq:Rinfinite}), where the denominator has a double zero associated with a singularity in the heat capacity. The Ricci scalar also diverges for $\a=0$, where the black hole is incompressible.
A correlation between singularities in response functions and singularities
in the Ricci scalar for thermodynamic state space is a generic feature 
of the geometry of black hole thermodynamic state space, \cite{RuppeinerReview}, and for black holes in particular, \cite{Mirza1}. 
Both Ricci scalars vanish on the hyperbola $\l\a^2=1$, except at the point $\l=\a^2=1$, since the denominators of both $R$ and $\widetilde R$ have double zeros there (equations (\ref{eq:Ricci_V}) and (\ref{eq:Ricci_VI}) in appendix \ref{app:KADS}).

The region of thermodynamic stability lies below the solid blue curve in the right-hand plots in figures \ref{fig:RicciW} and \ref{fig:RicciR} (the black hole is also unstable against the Hawking-Page phase transition for $\l<1$), the thermodynamically stable regime is isolated from the divergences in the heat capacity, except at the single point $\a^2=\l=1$. 
The Ruppeiner scalar is positive close to the line of the Hawking-Page phase transition, but can become negative for large pressure and entropy.

\section{Conclusions \label{sec:Conclusions}}

The thermodynamic geometry of asymptotically anti-de Sitter black holes 
in $D$ space-time dimensions has been examined in three special cases: electrically charged static black holes with fixed cosmological constant; electrically neutral singly spinning asymptotically AdS Myers-Perry black holes with fixed cosmological constant and electrically neutral singly spinning asymptotically AdS Myers-Perry black holes in an extended state
space in which the cosmological constant is included among the thermodynamic variables.  The first two cases
generalise previous studies in $D=4$ case to arbitrary dimension, while the last is a new direction in the geometry of black hole thermodynamics in which 
the black hole volume is interpreted as the thermodynamic variable conjugate to the positive pressure supplied by a negative cosmological constant.
The analysis is based on equation (\ref{eq:Weinhold-general}) which expresses the thermodynamic
metric in a manifestly general co-ordinate co-variant way, allowing any convenient co-ordinate system to be used for the calculations.

Our first observation extends the analysis of \cite{4D-RN-K-KN-metric} 
for 4-dimensional spherical black holes to $D$ space-time dimensions
and also to event horizons with positive ($k=+1$), negative ($k=-1$) and vanishing ($k=0$) curvature.
At fixed $\Lambda\leq 0$ the sign of the Weinhold curvature, derived from $M(S,Q,P)$ 
for a charged, non-rotating black hole in $D$ space-time dimensions, is determined by the topology of the event horizon: the Weinhold curvature
is positive for spherical event horizons and negative for hyperbolic horizons,
while flat event horizons give a flat Weinhold metric. 
This behaviour shows a correlation with thermodynamic stability, as only $k=1$ black holes can support a Hawking-Page phase transition.
In contrast the Ruppeiner metric is flat for $\Lambda=0$
but can be of either sign for $\Lambda <0$. 

For spherical event horizons the black hole can rotate and the
Weinhold metric on the 2-dimensional state space of a singly spinning electrically neutral black holes, derived
from $M(S,J,P)$ with $P$ fixed, is flat when $\Lambda=0$.
 
For asymptotically AdS black holes which are both charged and spinning there
is a critical point for fixed $Q$ and $J$ \cite{CCK}, but not for fixed $\Phi$
or fixed $\Omega$.  General considerations
suggest that the thermodynamic curvatures should diverge on spinodal 
curves, but the analysis in \S\ref{sec:AdSRN} and \S\ref{sec:MP} shows that the thermodynamic curvatures do not diverge on the singularities of $C_Q$ or $C_J$. 
Rather, due to cancellations
in $C_Q {\cal C}_T = C_\Phi {\cal C}_S$ and $C_J {\cal I}_T = C_\Omega {\cal I}_S$,
they diverge on the singularities of $C_\Phi$ and $C_\Omega$.  Contrary to expectations the thermodynamic curvatures do not diverge at the critical points. This unusual feature of
the geometry of the thermodynamic state space for black holes may be due to
the inhomogeneous scaling of the thermodynamic variables that makes 
the Smarr different from the integrated form of the 
Gibbs-Duhem relation.

For singly spinning electrically neutral black holes the thermodynamic state space can be enhanced to include the volume as a thermodynamic variable. The internal energy is a function of volume rather than pressure, $U(S,J,V)$ where a negative cosmological constant provides a positive pressure and 
$V=\left.\frac{\partial M}{\partial P}\right|_{S,J}$, with $M$ the black hole mass, is the thermodynamic volume.  Unlike the mass $M(S,J,P)$, which is the enthalpy in thermodynamics, the internal energy
$U(S,J,V)$ is a concave function of all its arguments
for thermodynamically stable systems and gives rise to a positive definite Weinhold and Ruppeiner metrics in the region of parameter space where the black hole is thermodynamically stable. 
The resulting Ricci scalars have been calculated and shown to
diverge for $J=0$, when the black hole is incompressible, and when the heat capacity at constant angular velocity and pressure, $C_{\Omega,P}$, diverges.  
Again the critical point is not visible in either the Weinhold or the Ruppeiner
curvature scalars.

\bigskip

\noindent Symbolic manipulations performed in this analysis were carried out using
Maple$^\copyright$.

\newpage

\appendix

\section{Thermodynamic curvature of charge black holes \label{app:RNADS}}

For non-rotating, asymptotically AdS black holes, the space-time metric is 
\begin{equation}
ds^2=-f(r) dt^2 +\frac {1}{f(r)}dr^2 + r^2 d^2\Omega_{(D-2)},
\end{equation}
where $d^2\Omega_{(D-2)}$ is the line-element on a unit $(D-2)$-dimensional 
sphere, torus or hyperbolic space for $k=1$, $0$ or $-1$ respectively,
and 
\begin{equation}
f(r) =k - \frac {2\mu}{r^{D-3}} + \frac{Q^2}{r^{2D-6}} + \lambda r^2.
\end{equation}
The mass and the entropy are
\begin{equation}
M= \frac{(D-2)\V\mu}{8\pi}
=\frac{(D-2)\V r_h^{D-3}
\left(k+\frac{Q^2}{r_h^{2D-6}}+r_h^2\lambda \right)}{16 \pi}
\end{equation}
and
\begin{equation}
S=\frac{\V r_h^{D-2}}{4}.
\end{equation}
$Q$ is the electric charge and $\lambda= -\frac{2\Lambda}{(D-1)(D-2)}$.

The entropy and the Hawking temperature are
\begin{equation}
S=\frac{\V}{4}r_h^{D-2}\label{eq:RN-entropy}\end{equation}
and
\begin{equation}
T=\frac{f'(r_h)}{4\pi}=\frac{(D-1)\lambda r_h^2 +(D-3)\left(k-\frac{Q^2}{r_h^{2D-6}} \right)}{4\pi r_h},
\end{equation}
where $\V$ is the volume of a unit $(D-2)$ sphere, torus or hyperbolic space
for $k=1$, $0$ or $-1$.\footnote{For $k=0$ and $-1$ the event horizon volume can be made finite by identifying points.}
For fixed $\Lambda<0$, the Weinhold metric is the $2\times 2$ Hessian matrix
obtained by differentiating $M(S,Q,P)$ with respect to $S$ and $Q$, 
keeping $P$ fixed,
\begin{equation}
g= \frac{\V\, \left( D-2 \right) }{8\pi }
\left( \begin {array}{cc} {r_h^{D-5} Z_Q(r_h,\lambda,Q)
}&-{\frac {  \left( D
-3 \right) Q}{ {{r_h}}^{D-2}}}\\\noalign{\medskip}
-{\frac { \left( D-3 \right) Q}{{{r_h}}^{D-2}}}&{
\frac {1}{{r_h}^{D-3} }}
\end {array} \right),
\end{equation}
where 
\begin{equation}
 Z_Q(r_h,\lambda,Q)=
\frac{1}{2}\left\{\left( D  -1\right){{\it r_h}}^{2}\lambda-
 (D  -3)k + (D-3)(2D-5)\frac {Q^2}{{r_h}^{2D-6 }}\right\}.
\end{equation}
$Z_Q(r_h,\lambda,Q)$ is of course related to the heat capacity at constant charge and pressure,
\begin{equation}
C_{Q,P}=
\frac{(D-2)\pi \V r_h^{D-1} T}
{2 Z_Q(r_h,Q,\lambda)}.
\end{equation}
The zero locus of $Z_Q(r_h,\lambda,Q)$ is the spinodal curve of $C_{Q,P}$.
This is however not visible in the determinant,
\begin{equation}
\det g = \frac{\V^2 (D-2)^2 Z_\Phi(r_h,Q,\lambda)}{64 \pi^2 r_h^2},
\end{equation}
where
\begin{equation}
Z_\Phi(r_h,Q,\lambda)=\frac{1}{2}\left\{(D-1)r_h^2\lambda -(D-3)\left(
k-\frac{Q^2}{r_h^{2D-6}}\right)\right\}.
\end{equation}
The determinant reflects the singularity structure of the heat capacity at constant electric potential, 
$\Phi=\left.\frac{\partial M}{\partial Q}\right|_{S,P}=\frac{(D-2)\V Q}{8\pi r_h^{D-3}}$, which is
\begin{equation}
C_{\Phi,P} =\frac{(D-2)\pi \V r_h^{D-1} T}
{2 Z_\Phi(r_h,Q,\lambda)}.
\end{equation}

The Weinhold scalar curvature is 
\begin{equation}
R=\frac{4 \pi k (D-3)^2}{(D-2)\V r_h^{D-3} Z_\Phi(r_h,Q,\lambda)^2}.
\end{equation}
For $k=1$ the singularities in the curvature match those of the heat capacity
of the black hole at constant pressure, $C_{\Phi,P}$.  For $k < 0$, $C_{\Phi,P}$
and $R$ are finite and $R$ is negative.

The Weinhold curvature scalar can be succinctly written in terms of the electrical capacitance,
\begin{equation}
{\cal C}_{S,P}=\left.\frac{\partial Q}{\partial \Phi}\right|_{S,P}=
\frac{8\pi r_h^{D-3}}{(D-2)\V},
\end{equation}
and the entropy (\ref{eq:RN-entropy}) as
\begin{equation} R=  \frac{ (D-3)^2}{(D-2)}\frac{\pi r_h k}{S Z_\Phi^2}.
\end{equation}


The Ruppeiner metric on the other hand, based on the Hessian matrix of $S(M,Q,P)$ with a 2-dimensional state space
consisting of $(M,Q)$ and $P$ fixed, has scalar curvature
\begin{equation}
\widetilde R=\frac{4(D-1)\lambda\bigl(3\pi T - (D-1) r_h\lambda \bigr)
\widetilde  F(\lambda,r_h,Q)  C_{\Phi,P}^2} 
{\bigl\{(D-2) \V r_h^{D-2} \pi T\bigr\}^3}
\end{equation}
where $\widetilde F(\lambda,r_h,Q)$ is a linear function of $\lambda$,
\begin{equation}
\widetilde F(\lambda,r_h,Q)=(D-1)(D-2) r_h^2\lambda + 
(D-3)\left\{(D-4)k +(D-2)\frac{Q^2}{r_h^{2D-6}}\right\},\label{eq:F_tilde}
\end{equation}
which is positive for $k=1$.

\section{Metric and Ricci scalar for rotating black holes \label{app:KADS}}

A singly spinning asymptotically AdS Myers-Perry black hole,
in $D>3$ dimensions,
has the line element \cite{HawkingTaylorRobinson}
\bea
ds^2 &=&
 -\frac{\Delta}{\rho^2} \left( dt -\frac{a}{\Xi} \sin^2\theta d\phi \right)^2
+\frac{\rho^2}{\Delta}dr^2
+\frac{\rho^2}{\Delta_\theta}d\theta^2 \nonumber \\
&& \qquad\qquad + \frac{\Delta_\theta \sin^2\theta}{\rho^2}
\left(a dt - \frac{r^2+a^2}{\Xi}d\phi \right)^2 + r^2 \cos^2\theta 
d\Omega_{(D-2)},\nonumber
\eea
where
\bea \Delta&=&(r^2 + a^2)\left(1+ \lambda r^2 \right) -\frac{2\mu}{r^{D-5}},\nonumber\\
\Delta_\theta&=&1-\lambda a^2 \cos^2\theta, \nonumber \\
\rho^2&=&r^2+ a^2\cos^2\theta, \nonumber\\
\Xi&=&1-\lambda a^2.\nonumber \eea
$\mu$, $a$ and $\lambda$ are geometric parameters related to mass, rotation and cosmological constant respectively (as before, $\lambda=-\frac{2\Lambda}{(D-1)(D-2)})$. 
There is an event horizon at the largest root, $r_h$, of
\[ \Delta(r_h)=0 \qquad \Rightarrow \qquad \mu = \frac 1 2 r_h^{D-5}(r_h^2 + a^2)\left(1+ \lambda r_h^2 \right).\]

The Hawking temperature is 
\beq
T=\frac{(D-3)(1+\lambda a^2)r_h^2  +(D-1)\lambda r_h^4  +(D-5)a^2}{ 4 \pi r_h(r_h^2 + a^2)},
\eeq 
for $D>4$ this is a positive function for any $a$.  The condition that $a^2\ge 0$
forces $T$ to lie in the range
\[(D-3)\lambda r_h^2 +(D-5) \le  4\pi r_h T \le (D-1)\lambda r_h^2 +(D-3).\]

When $\Lambda$ is fixed the Weinhold metric derived from $M(S,J,P)$ with fixed $P$ is, in $(r_h,a)$ co-ordinates,
\bea
g&=&\nonumber\\
&& \kern -70pt \frac{\V r_h^{D-5}}{16\pi (r_h^2 + a^2)}
\left(\begin{matrix}
 \frac{p_1(r_h,\lambda,a)}{ (1-\lambda a^2)} & 
-\frac {2 a\left( 1- \lambda^2r_h^4\right) \left((D-1)r_h^2+ (D -5)\,{a}^{2} \right)}{r_h\left( 1-\lambda\,{a}^{2} \right)^2} \\
-\frac {2 a\left( 1- \lambda^2r_h^4\right) \left((D-1)r_h^2+ (D -5)\,{a}^{2} \right)}{r_h\left( 1-\lambda\,{a}^{2} \right)^2}  & 
\frac{2 (1+\lambda r_h^2)^2
\bigl[ (3-a^2)\lambda a^2 + r_h^2-3 a^2 \bigr]}{(1-\lambda a^2)^3} \\
\end{matrix}\right), \nonumber
\eea
where $p_1(r_h,\lambda,a)$ in the top-left entry is
\bea
p_1(r_h\lambda,a)&=&\lambda\bigl[(D-3)(D-4)a^4 + 2(D^2-5D+3) a^2 r_h^2 +(D-1)(D-2)r_h^4\bigr] \nonumber \\
&& -(D-4)(D-5)a^4 -2(D^2-7D+9)a^2r_h^2-(D-2)(D-3)r_h^4.\nonumber
\eea

The Ricci scalar arising from this Weinhold metric is,
in terms of dimensionless variables $\l=\lambda r_h^2$ and $\a=\frac{a}{r_h}$ ,
\beq
R=\frac{16\pi \l(1-\l\a^2)F_1(\l,\a)}{\V r_h^{D-3}(1+\a^2) \{D-2 +(D-4)\l\a^2  \}^2 Z_\Omega^2(\l,\a)},
\eeq
where 
\beq
Z_\Omega(\l,\a)=(D-3)(1+\l\a^2)  -(D-1)\l -(D-5)\a^2,\label{eq:Gminus}
\eeq
and $F_1(\l,\a)$ is linear in $\l$,
\beq
F_1(\l,\a)= A_1(\a) \a^2 \l + B_1(\a), \label{eq:F1}
\eeq
with
\bea A_1(\a)&=&  (D-5)^2 \a^4 \bigl\{ (D-4)\a^2+ (D-6)\bigr\} \nonumber \\
&& \hskip 80pt  -(D-1)(D^2-17D +48) \a^2  -(D-1)^2(D-2), \nonumber \\ 
B_1(\a)&=& -(D-5)\a^4\bigl\{(D-4)(D-5)\a^2 +(D^2 +5D -18)\bigr\} \nonumber \\
&& \hskip 120pt+(D-1)^2\bigl\{ D\a^2+(D-2)\bigr\}.\nonumber
\eea

The zeros of $Z_\Omega$ reflect singularities in the heat capacity,
in this instance the heat capacity at constant angular velocity and pressure,
\begin{equation}
C_{\Omega,P}=-\frac{4\pi r_h T S \bigl\{D-2-(D-4)\a^2 \bigr\}}
{Z_\Omega(\l,\a)}.
\end{equation}
The heat capacity at constant angular momentum and pressure has a rather
different structure,
\begin{equation}
C_{J,P}=-\frac{4\pi r_h T S (1+\a^2)^2 \bigl\{D-2+(D-4)\lambda\a^2 \bigr\}}{Z_J(\l,\a)}
\end{equation}
with
\begin{eqnarray*}
Z_J(\l,\a)&=&-\l^2\a^2\bigl\{(D-3)\a^4 +6 \a^2 +3(D-1)\bigr\} \\ 
&& \quad +\l\bigl\{(D-5)\a^6 + (5D-33)\a^4 -(5D+3)\a^2 -(D-1)\bigr\} \\
&& \quad\quad +3(D-5)\a^4 -6\a^2 +D-3.\label{eq:ZJ}
\end{eqnarray*}

The isentropic moment of inertia at constant pressure (analogous to $\kappa_S$ for a gas
and evaluated in \cite{Stability}) is
\begin{equation}
{\cal I}_{S,P}= 
\left.\frac{\partial J}{\partial \Omega}\right|_{S,P}=
\frac{r_h S (1+a^2)^2 \bigl\{D-2+(D-4)\lambda\a^2 \bigr\} }{2 \pi (1-\l \a^2)^2\bigl\{ D-2 -(D-4)\a^2\bigr\}}.\label{eq:I_S}
\end{equation}
The isothermal moment of inertial tensor ${\cal I}_{T,P}$ can be obtained from the identity $C_{\Omega,P}{\cal I}_{S,P}=C_{J,P}{\cal I}_{T,P}$,
\begin{equation}
{\cal I}_{T,P}=\frac{r_h S Z_J(\l,\a)}{2\pi (1-\l \a^2)^2 Z_\Omega(\l,\a)}.\label{eq:I_T}
\end{equation}

The Ruppeiner metric has Ricci scalar
\beq
\widetilde R= -\frac{(D-3)(1-\l^2 \a^4)
\widetilde F_1(\l,\a)\widetilde F_2(\l,\a)}{\pi \V r_h^{D-1} T (1+\a^2)^2 \bigl\{D-2+(D-4)\l\a^2 \bigr\} Z_\Omega^2(\l,\a)}\label{app:R_tilde_fixed_P}
\eeq
where $\widetilde F_1(\l,\a)$ is linear in $\lambda$ and $\a^2$
while $\widetilde F_2(\l,\a)$ is quadratic. Explicitly
\bea
\widetilde F_1(\l,\a)&=&(D-3)(1+\l\a^2)-3\bigl\{(D-1)\l +(D-5)\a^2\bigr\}
\label{eq:F1_tilde}\\
\widetilde F_2(\l,\a)&=&  \widetilde A_2(\a)\l^2 +  \widetilde B_2(\a)\l +  \widetilde C_2(\a),\label{eq:F2_tilde}
\eea
with
\bea
 \widetilde A_2(\a)&=&(D-4)\bigl\{ (D-3)\a^2 -(D-1)\bigr\}\a^2,\nonumber\\
 \widetilde B_2(\a)&=& (D-5)(D-6)\a^4-2(D^2-6D+1)\a^2 +D(D-1),\nonumber\\
 \widetilde C_2(\a)&=& (D-2)\bigl\{(D-3) - (D-5)\a^2\bigr\}.\nonumber
\eea
Note that the only singularities 
in $R$ or $\widetilde R$ are those associated with the spinodal curve of 
$C_{\Omega,P}$, $Z_\Omega=0$.

When $\l=0$ (\ref{app:R_tilde_fixed_P}) reduces to 
\beq
\widetilde R= -\frac{4(D-3)\bigl\{D-3 -3(D-5)\a^2 \bigr\}}
{\V r_h^{D-2}(1+\a^2)\bigl\{(D-3)^2-(D-5)^2\a^4 \bigr\}},\label{Ruppeiner_Scalar}
\eeq
which is the result quoted in \cite{D-RN-Kerr}.

There is a critical point on the spinodal curve for $C_{J,P}$.
This was first
found for $D=4$ in \cite{CCK}, where the critical values are $\l_{crit}=0.2105$ and 
$\a_{crit}=0.1795$, and is present in any space-dimension greater
than three. In 5-D for example $\l_{crit}=0.3569$ and  
$\a_{crit}=0.1802$.  
There is no critical point visible in $C_{\Omega,P}$, in fact $Z_\Omega>0$ and $C_\Omega<0$ at the critical point and the critical point is actually unstable if $J$
is not fixed.  The
curvature scalars diverge on the spinodal curve for $C_{\Omega,P}$,
not that of $C_{J,P}$, and the critical point is not visible in the thermodynamic 
curvature.

Allowing $P$ to vary requires using $V$ as a thermodynamic variable in order
to ensure a positive definite Weinhold metric, associated with  $U(S,J,V)$, in regions of the state space where the black hole is thermodynamically stable. 
The Weinhold metric following from (\ref{eq:Weinhold-general}), (\ref{eq:SJV}), (\ref{eq:TOP}) and (\ref{eq:MU}) is
\bea g&=& \frac{\V r_h^{D-5}}{16\pi} \times  \label{eq:WMP} \\
&&\kern -1.5cm 
\left[ \begin{array}{ccc} 
\frac{(1+\a^2)p_2(\l,\a)}{(1-\l\a^2)^2}
& -\frac{2(D-2)r_h(1-\l^2)\a}{(1-\l\a^2)^2} & 
-\frac{2 r_h [(D-1)-(D-3)\l](1+\a^2)\a^2}{(1-\l\a^2)^2} \\ 
 -\frac{2(D-2)r_h(1-\l^2)\a}{(1-\l\a^2)^2} &
\frac{2 r_h^2 (1+\l)^2 (\a^2(3-\a^2)\l+1-3\a^2 )}{(1+\a^2)(1-\l\a^2)^3}  &
\frac{r_h^2(1+\l)(\l\a^2 + 1-2\a^2)\a}{(1-\l\a^2)^3} \\
-\frac{2 r_h [(D-1)-(D-3)\l](1+\a^2)\a^2}{(1-\l\a^2)^2}  &
\frac{r_h^2(1+\l)(\l\a^2 + 1-2\a^2)\a}{(1-\l\a^2)^3}  &
\frac{2 r_h^2(1-\a^4)\a^2}{(1-\l\a^2)^3} \\
\end{array}\right]\nonumber \eea
where the top left entry involves
\[p_2(\l,\a)=-D(D-3)\a^2\l^2 +
\bigl\{(D^2-5D+8)\a^2 + (D-1)(D-2) \bigr\}\l -(D-2)(D-3). \]
The determinant of the Weinhold metric, in $(r_h,\l,\a)$ co-ordinates, is 
\beq \det g = -\frac{\V^3 (D-3) r_h^{3D-11} (1+\l)^3 (1+\a^2) \a^4 Z_\Omega(\l,\a)}{2048 \pi^3 (1-\l\a^2)^6},\label{ew:detg}\eeq
which, for $\l>0$, vanishes when $\a=0$ and when
\begin{equation}
\l :=\l_\infty =\frac{(D-3)-(D-5)\a^2}{(D-1)-(D-3)\a^2},
\end{equation}
and diverges for
\[\l\a^2=1. \]
The zeros of $\det g$ are genuine curvature singularities and are reflected
in the Ricci scalars below, indeed these reflect singularities in the response
functions: the black hole is incompressible for $\a^2=0$ and the heat capacity
at constant angular velocity,
$C_{\Omega,P}$, diverges on $\l_\infty$ where $Z_\Omega=0$. 
Both Ricci scalars vanish at the limit of thermodynamic state space where $\l\a^2=1$ and extensive quantities diverge.
 
 The Ricci scalar following from the Weinhold metric (\ref{eq:WMP}) can be written as a ratio of two
polynomials in the dimensionless variables $(\l,\a^2)$
\begin{equation} R=\frac{16\pi}{(D-3) \V r_h^{D-3}} \frac{(1-\l \a^2)F_2(\l,\a^2)}{\a^4 ( 1+{\a}^{2})\bigl( 1+\l\bigr)^{2} Z_\Omega^2(\l,\a),},
\label{eq:Ricci_V}\end{equation}
where $F_2$ is quadratic in $\l$ and quartic in $\a^2$. Explicitly
\begin{equation}
F_2(\l,\a^2)=A_2(\a^2)\l^2 + B_2(\a^2)\l + C_2(\a^2) \label{eq:F2}
\end{equation}
with
\bea
A_2(\a^2)&=& {\a}^{2}\left[ \left( D-3 \right)  \left( 7\,D-27 \right) {\a}^{6}-
\left(5\,{D}^{2}-10\,D  - 19\right) {\a}^{4}\right. \nonumber \\
&& \qquad \qquad \left.-\left( D-1 \right)  ( 3\,D-13) {\a}^{2} 
+\left( D-1 \right) ^{2} \right], \nonumber \\
&&\nonumber\\
B_2(\a^2)&=& 
- (D-2)(D-5)(2\,D-7)   {\a}^{8}
+ ({D}^{3}-9\,{D}^{2}+20\,D -4) {\a}^{6}\nonumber\\
&&
+ ( 3\,{D}^{3}-17\,{D}^{2}+22\,D+12) {\a}^{4}
-( D-1 ) ( {D}^{2}-8 ) {\a}^{2}- \left( D-1 \right)^2  \left( D-2 \right), \nonumber\\
&&\nonumber\\
C_2(\a^2)&=&
\left( D-3 \right)  \left( D-5 \right) ^{2}{\a}^{8}+ \left( D-5 \right)  \left( {D}^{2}-11\,D+19 \right) {\a}^{6}\nonumber \\
&& \quad - \left(3\,{D}^{3}-28\,{D}^{2} +70\,D-61 \right) {\a}^
{4} - \left({D}^{3}-12\,{D}^{2} +30\,D-23 \right) {\a}^{2}\nonumber \\
&&\hskip 3cm + ( D-1 )  ( D-2 ) ( 2\,D-5).\nonumber
\eea
The only singularities of $R$ are those associated with the zeros of $Z_\Omega$.

The Ricci scalar for the Ruppeiner metric conformal to (\ref{eq:WMP}) is more
complicated,
\beq
\widetilde R = \frac{(1-\l\a^2)\widetilde F_3(\l,\a)}
{4\pi (D-3)\V r_h^{D-1} \a^4 (1+\a^2)^3 (1+\l)^2 T  Z_\Omega^2},
\label{eq:Ricci_VI}\eeq
 where $\widetilde F_3(\l,a)$ is a quintic polynomial in $\l$ and sixth order in $\a^2$. Again the only singularities are associated with the zeros of $Z_\Omega$.

\end{document}